\documentstyle[amssymb,graphicx,multicol,epsfig,prl,aps]{revtex}

\begin{document}
\draft
 
\title{Cracks and Crazes: On calculating the macroscopic fracture energy of
glassy polymers from molecular simulations}

\author{J\"org Rottler, Sandra Barsky, and Mark~O.~Robbins}

\address{Department of Physics and Astronomy, The Johns Hopkins
University, 3400 N.~Charles Street, Baltimore, MD 21218}

\date{\today}

\maketitle

\begin{abstract} 
We combine molecular dynamics simulations of deformation at the
submicron scale with a simple continuum fracture mechanics model for
the onset of crack propagation to calculate the macroscopic fracture
energy of amorphous glassy polymers.  Key ingredients in this
multiscale approach are the elastic properties of polymer crazes and
the stress at which craze fibrils fail through chain pullout or
scission. Our results are in quantitative agreement with dimensionless
ratios that describe experimental polymers and their variation with
temperature, polymer length and polymer rigidity.
\end{abstract}

\pacs{PACS numbers: 82.35.Lr, 83.60.Uv, 62.20.Mk}

\begin{multicols}{2}

\narrowtext

Understanding the molecular origins of macroscopic mechanical
properties such as the fracture energy $G_c$ is a fundamental
scientific challenge.  In tough materials, the work $G_c$ required to
propagate a crack through a unit area is orders of magnitude higher
than the lower bound provided by the
equilibrium interfacial free energy $G_{\rm eq}$ of the crack
surfaces.  Efforts to calculate this large increase in fracture
energy have been frustrated,
because phenomena on many length scales must be
treated simultaneously \cite{MRS2001}.  In both amorphous and
crystalline materials, the fracture energy depends on processes that range
from breaking of atomic bonds to formation of defect structures on
micron and larger scales.

In this Letter, we present a multiscale approach that allows us to
calculate the plane strain fracture energy of an important class of unfilled
amorphous polymer glasses. Experiments
\cite{Haward1997,Brown1991,Kramer1990} show that under tensile loading
the fracture energy of materials such as polystyrene (PS) and
polymethylmethacrylate (PMMA) is mainly due to the formation of an
intriguing craze structure in a ``process zone'' around the crack tip
(Fig.~\ref{activezone-fig}). This results in a large increase in
fracture energy,
$G_c/G_{eq} \sim 10^3 - 10^4$, that is essential to the use of
these materials as adhesives, packaging materials and windows
\cite{Wool1995,Pocius1997,Haward1997}.

In the craze, $\sim 0.5$nm diameter polymers are bundled into an
intricate network of $\sim 10$nm diameter polymers that extends $\sim
10\mu$m on either side of the $\sim $ mm crack and $\sim 100\mu$m
ahead of the crack tip.  Molecular level simulations of regions with
linear dimensions of mm or even $\mu$m are not feasible.  They would
also be inefficient, since most regions near the crack are homogeneous
enough to be treated with continuum mechanics
\cite{Brown1991,Bernstein2000}. Here, we combine the two approaches
using molecular simulations of representative volume elements (see
Fig.~\ref{activezone-fig}) to provide information about craze
formation, deformation and failure that is needed to construct a
continuum fracture mechanics model.

One advantage of studying polymeric systems is that many dimensionless
ratios are independent of the specific chemistry of the molecules.
For this reason we consider a bead-spring model that has been shown to
provide a realistic description of polymer behavior
\cite{Puetz2000,Baljon2001,Faller2000,Sides2001}. Each linear polymer
contains $N$ beads of mass $m$.  Van der Waals interactions between
beads separated by a distance $r$ are modeled with a truncated
Lennard-Jones potential: $V_{\rm
LJ}(r)=4u_0\left[(a/r)^{12}-(a/r)^{6}-(a/r_c)^{12}+(a/r_c)^{6}\right]$
for $r\le r_c=1.5\,a$, where $u_0$ and $a$ are characteristic energy
and length scales.  A simple analytic potential, $V_{\rm
br}(r)=-k_1(r-r_c)^3(r-R_1)$, is used for the covalent bonds between
adjacent beads along each chain.  The constants $k_1$ and $R_1$ are
adjusted to fix the equilibrium bond length \cite{Puetz2000},
$0.96a$, and the ratio of the forces at which covalent and van
der Waals bonds break.  We find that this ratio is the only important
parameter in the covalent potential and set it to 100 based on data
for real polymers \cite{Sides2001,Stevens2001}. The polymer rigidity
and entanglement length $N_e$ are varied by introducing local
bond-bending forces \cite{Faller2000} with a potential
$V_{B}=b\sum_{i=2}^{N-1}\left(1-\frac{(\vec{r}_{i-1}-\vec{r}_{i})\cdot
(\vec{r}_{i}-\vec{r}_{i+1})}{|(\vec{r}_{i-1}-\vec{r}_{i})||(\vec{r}_{i}-
\vec{r}_{i+1})|}\right)$ along the backbone. $\vec{r}_{i}$ denotes the
position of the $i$th bead along the chain, and $b$ characterizes the
stiffness. Two limiting cases of fully flexible $(N_e\sim 60-70$
beads, $b=0u_0$) and semiflexible $(N_e\sim 30$ beads, $b=1.5u_0$)
chains are considered here.  The chain length is varied from $N=64$
beads to $1024$ beads.

We first show that our model captures the essential experimental
features of craze formation (Fig.~\ref{activezone-fig}A).  The
simulation cell has periodic boundary conditions and is initially a
cube of size $L$. The length along one direction $L_3$ is increased at
constant rate, while the other dimensions of the cell are held fixed.
Fig.~\ref{smax-fig}(a) shows typical results for the stress $\sigma_3$
along the stretching direction as a function of the elongation
$L_3/L$. In all cases, there is an initial peak at small strains,
where the material yields by cavitation \cite{Rottler2001,Baljon1996}.
As in experiment \cite{Kramer1990}, this peak is followed by a long
plateau at a constant stress $S$.  During this plateau, deformation is
localized in a narrow ``active zone'' at the boundary of the growing
craze network (Fig.~\ref{activezone-fig}A). 

%****************************************************************
\begin{center}
\begin{minipage}{18cm}
\begin{center}
\begin{figure}[hbt]
\epsfig{file=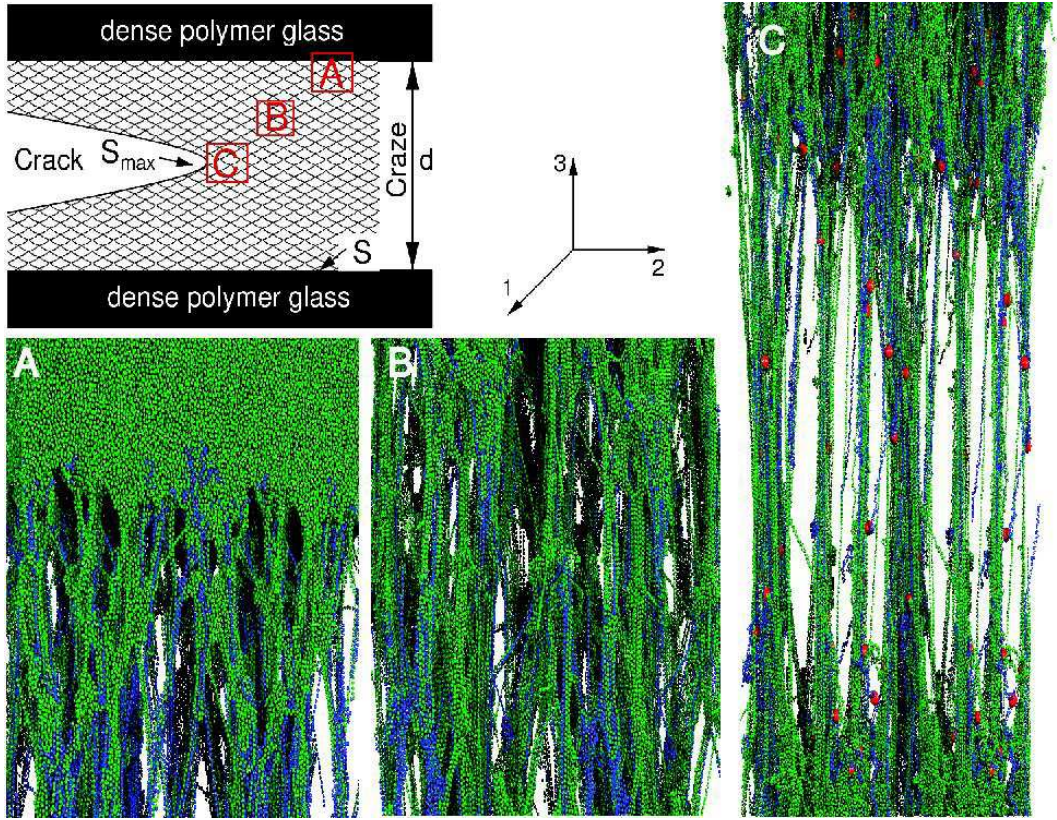,width=14cm}
\vspace*{0.1cm}
\caption{Some length scales involved in the fracture of glassy
polymers.  A $\sim$mm long crack penetrates the material. A craze of
typical width $d =10-50\mu$m and length $l \sim 10 d$ forms in the
``process zone'' in front of the crack tip (top left). Craze formation
(A), deformation (B) and failure (C) are studied with molecular
dynamics simulations of $\sim $100 nm-sized volume elements.  A: craze
fibrils emerge from the dense polymer glass (top) within a narrow
``active zone'', B: fully developed craze structure, C: failure of the
craze immediately in front of the crack tip. Individual beads of the
polymer chains are shown as small ellipsoids. Chains carrying the
largest tension are colored blue and broken bonds in C are colored
red. The width of each image is 128 bead diameters. The fibril spacing
$D_0$ is $\sim $10 bead diameters for this figure, but depends on
temperature, chain rigidity and other parameters.}
\label{activezone-fig}
\end{figure}
\end{center}
\end{minipage}
\end{center}
\vspace{0.01cm}
%****************************************************************

$S$ represents the stress needed to draw fibrils out of the dense
regions adjacent to the craze \cite{Baljon2001}.  This steady state
drawing process increases the volume occupied by the polymer by a
constant factor called the extension ratio $\lambda$.  When $L_3/L$
reaches $\lambda$, the entire system has evolved into a craze, and the
stress (Fig.~\ref{smax-fig}(a)) begins to rise.

It is evident from Fig.~\ref{smax-fig}(a) that $\lambda$ is strongly
dependent on chain rigidity and thus the entanglement length $N_e$.
We find that $\lambda$ decreases from about 6.1 for flexible chains to
3.6 for semiflexible chains.  As in experiments, these values are
quantitatively consistent with a simple model that assumes
entanglements act like permanent chemical crosslinks
\cite{Kramer1990}. During crazing, segments between entanglements are
expanded from their equilibrium random-walk configurations to nearly
straight lines (Fig.~\ref{activezone-fig}B).

We consider the common case where the dominant contribution to the
fracture energy is the work needed to craze material in the process
zone ahead of the crack tip\cite{Haward1997}. As the crack advances,
each region is expanded \\
\vspace*{14.5cm} 

at the constant plateau stress $S$. In steady state, advancing the
crack over an area $A$ has the net effect of expanding a region of
this area at constant stress from its initial width to the final width
$d$ at which the craze cracks.  Thus $G_c=S(d-d/\lambda)$
\cite{dugdale-comm}.  After normalizing by the lower bound for the
fracture energy provided by the interfacial free energy change $G_{\rm
eq}=2\gamma$, the fracture energy can be written in dimensionless form
as
\begin{equation}
\label{gc-eq}
\frac{G_c}{G_{\rm eq}}=\frac{SD_0}{2\gamma}\frac{d}{D_0}(1-1/\lambda).
\end{equation}
Eq.~(\ref{gc-eq}) shows that $G_c$ is primarily limited by the craze
width $d$. Unlike the other quantities in Eq.~(\ref{gc-eq}), $d$
cannot be obtained directly from MD simulations.  However, a minimal
continuum model proposed by Brown \cite{Brown1991} allows us to
calculate $d$.

Brown pointed out that although there is a constant plateau stress $S$
on the craze boundary, a stress concentration occurs near the crack
tip (see Fig.~\ref{activezone-fig}). He formulated a fracture
criterion by equating the stress at the crack tip to the maximum
stress $S_{\rm max}$ that the craze fibrils can withstand. The stress
at a distance $r$ from a crack tip in a continuous elastic medium
diverges as $r^{-1/2}$.  This divergence is cut off at the
characteristic fibril spacing $D_0$, below which the material can no
longer be treated as a homogeneous elastic medium.  Since the stress
varies from $S_{\rm max}$ to $S$ as $r$ varies from $D_0$ to $d/2$,
$S_{\rm max}\propto S\left(d/D_0\right)^{1/2}$.  Solving the linear
fracture mechanics problem within the craze yields \cite{Paris1965}
\begin{equation}
\frac{d}{D_0}=4\pi \kappa \left(\frac{S_{\rm max}}{S}\right)^2,
\label{fracture-eq}
\end{equation}
where $S_{\rm max}$ denotes the maximum stress that the craze fibrils
can withstand, and the prefactor $\kappa$ depends on the anisotropic
elastic constants $c_{ij}$ of the craze network
(Eq.~(\ref{kappa-eq})). Neither the elastic properties of the craze
network nor $S_{\rm max}$ are easily obtained from experiments.
However, we can calculate both from MD simulations of regions B and C
in Fig.~\ref{activezone-fig} and thereby provide the key ingredients
for calculating the fracture energy of glassy polymers from
Eqs.~(\ref{gc-eq}) and (\ref{fracture-eq}).

Elastic constants were calculated by applying small ($\leq 0.5$\%)
step strains to fully developed crazes (Fig. 1B) at two different
elongations $L_3/L$ and measuring the change in stress.  Table I shows
key ratios for flexible and semiflexible chains at two representative
temperatures $T=0.1\, u_0/k_B$ and $T=0.01\, u_0/k_B$. Both are well
below the glass transition temperature $T_g\approx 0.35\,u_0/k_B$ of
the bead-spring model.  As can be expected from the highly oriented
structure of the craze network (see Fig.~1), $c_{33}$ is always much
bigger than the other elastic constants, which are all of the same
order. The prefactor $\kappa$ in Eq.~(\ref{fracture-eq}) is given by
\cite{Paris1965}
\begin{equation}
\kappa^2=\frac{(1-C_2)+(c_{33}/2c_{44})(1-C_1)}{2(1-C_1)^2},
\label{kappa-eq}
\end{equation}
where $C_1\equiv 2C_2c_{13}/c_{33}$ and $C_2\equiv
c_{13}/(c_{11}+c_{12})$.  Inserting the elastic constants, we obtain
values for $\kappa$ between 2.0 and 2.8 for flexible chains and
between 1.1 and 1.7 for semiflexible chains. The crazes with higher
elongations always have a lower value of $\kappa$.

A simple approximate expression $\kappa\approx\sqrt{c_{33}/4c_{44}}$
can be obtained by noting that $c_{33}\gg c_{13}$, and thus $C_1\sim
0$ and $C_2\sim 1$. Table I also shows that this is an accurate
approximation for all practical purposes.  This simple expression
shows clearly that the ability of crazes to resist shear $(c_{44}>0)$
limits their fracture energy. As first pointed out by Brown
\cite{Brown1991}, the absence of lateral stress transfer would lead to
$\kappa \rightarrow \infty$ and thus to an infinite $G_c$.

To determine the stress $S_{\rm max}$ at which fibrils break, we
continue straining the fully developed craze until it fails
(Fig.~\ref{activezone-fig}C).  Although all chains that are long
enough to form stable crazes ($N/N_e \gtrsim 2 $ \cite{Baljon2001}) show
the same plateau stress and extension ratio, their crazes exhibit very
different behavior for $L_3/L > \lambda$ (Fig. 2(a)).  Short chains of
length $N/N_e=2$ easily pull free from the topological constraints
imposed by entanglements, and the stress drops monotonically.  As $N$
increases, the force needed to pull chains free from entanglements
along a failure plane rises, and there is a corresponding increase in
$S_{\rm max}$.  The failure mechanism changes when the force needed to
disentangle the chains reaches the breaking force for covalent bonds.
At this point the forces along the chains and $S_{\rm max}$ both
saturate due to chain scission.
%****************************************************************
\begin{figure}[tb]
\epsfig{file=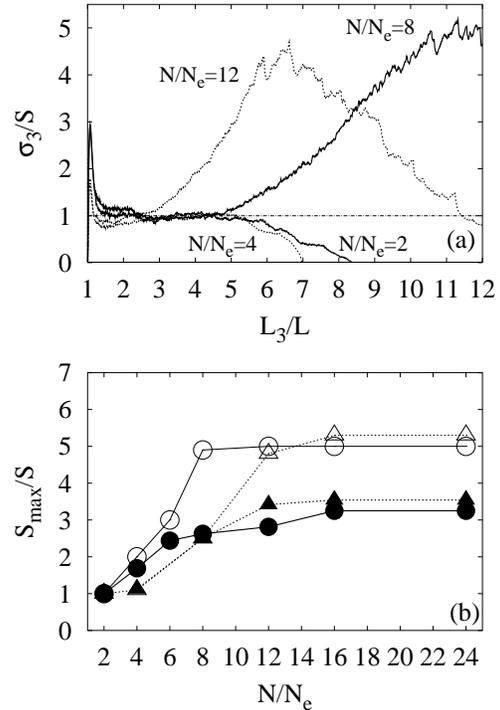,width=7cm}
\caption{(a) Normalized stress $\sigma_3/S$ during craze growth at
$T=0.3\,u_0/k_B$ for flexible (solid lines) and semiflexible (dotted
lines) chains of the indicated length in systems composed of 32768
beads. Here we take $N_e=32$ for the semiflexible and $N_e=64$ for the
flexible chains. The value of $S_{\rm max}/S$ is obtained from the
maximum height of the curves. Panel (b) summarizes $S_{\rm max}/S$ for
the flexible ($\circ$) as well as semiflexible ($\triangle$) chains at
$T=0.3\,u_0/k_B$ (open symbols) and $T=0.1\,u_0/k_B$ (filled symbols)
as a function of chain length.}
\label{smax-fig}
\end{figure}
%****************************************************************
Fig.~\ref{smax-fig}(b) summarizes our results for $S_{\rm max}/S$ as a
function of chain length, temperature and flexibility.  As $N/N_e$
rises above 2, $S_{\rm max}/S$ rises rapidly and then saturates due to
the change in failure mechanism from chain pullout to chain scission.
Saturation occurs for $N/N_e$ between 8 and 16. The limiting value of
$S_{max}/S$ lies between 3.4-3.8 for $T\leq 0.1u_0/k_B$ and 5.0-5.3 for
$T=0.3u_0/k_B$ and increases slightly with chain rigidity. In general,
more chain scission is observed for chains with higher rigidity and at
lower temperatures.

We are now in a position to evaluate Eq.~(\ref{fracture-eq}) and
compare our results to experimental values. For flexible chains,
$\kappa=2.0-2.8$ and $S_{\rm max}/S=3.4-5.0$, yielding $d/D_0$ between
290 - 890. Values for semiflexible chains give $d/D_0$ between 200 -
600. Typical craze widths $d$ observed in experiments range between
$3-20\,{\rm \mu m}$, whereas characteristic fibril spacings $D_0$ have
values between $20-30\,{\rm nm}$.  Thus the range of experimental
values for $d/D_0= 100 - 1000$ overlaps well with our results.

In addition to values quoted above, calculating the fracture energy
from Eq.~(\ref{gc-eq}) requires values for the plateau
stress $S$, mean fibril spacing $D_0$, and surface tension
$\gamma$. Typical values from our simulations are $S=0.5 -
1.4\,u_0/a^3$, $D_0=10 - 14\,a$, and $\gamma=0.6-1.0\,u_0/a^2$. With
these values, we arrive at our final result $G_c/G_{\rm eq}=1300 -
4300$ (flexible polymers) and $G_c/G_{\rm eq}=1200 - 3500$
(semiflexible polymers). Within this range, $G_c/G_{\rm eq}$ tends to
drop with increasing chain rigidity and decreasing temperature
$T$. This trend is also found in real adhesive joints, where the
fracture energy generally decreases with decreasing temperature
\cite{Wool1995}.

Our simulations agree with experimental observations in greater
detail. Sha {\it et. al.} \cite{Sha1995} have compiled values of $G_c$
for PS and PMMA as a function of polymer molecular weight
$M_w$. Neither polymer shows a large fracture energy when $M_w$ is
less than twice $M_e$. As in our simulations (Fig. 2(b)), the fracture energy
rises rapidly as $M_w/M_e$ rises above 2 and then saturates around 10
$M_e$. The limiting values of $G_c/G_{eq}$ at large $M_w/M_e$,
2500 (PMMA) and 5000 (PS), are comparable to our predicted values.

In conclusion, we have demonstrated that supplementing a simple
continuum model with constitutive relations from molecular simulations
can provide quantitative predictions for key material parameters such
as the fracture energy. This approach can be further developed by
using chemically detailed interaction potentials for specific polymers
in the molecular simulations.
For example, one might expect that a realistic treatment of side groups
could increase the friction between polymers during craze formation,
and increase the likelihood of chain scission.
A more detailed finite element model
for crack propagation could also be used, as in recent work
\cite{Tijssens2000,Socrate2001} that assumed simple constitutive
relations for craze widening and breakdown.  Although one might hope
to include all length scales simultaneously in a hybrid calculation,
this is complicated by the rapid increase in the relevant time scale
with increasing length scale \cite{Bernstein2000}.  Finally, it should
be noted that macroscopic cracks can contain an ensemble of crazes
with characteristic sizes and spacings and that other processes such
as shear banding can contribute to the fracture energy, depending
on loading conditions and materials. These issues should provide
fruitful topics for future work.

This work was supported by the Semiconductor Research Corporation and
National Science Foundation Grant 0083286. We thank H.~R.~Brown,
E.~J.~Kramer, and C.~Denniston for stimulating discussions.

\end{multicols}

%****************************************************************
\begin{table}
\caption{Elastic properties of model crazes composed of 262 144 beads
with flexible (fl.) and semiflexible (sfl.) chains at two elongations
$L_3/L$. Uncertainties in $c_{ij}$ are about 10\%. Here $N=256$, but
the results do not depend on $N$ for $N>2N_e$. They are also
insensitive to the covalent bond potential, since strain is
accommodated by the weaker van der Waals bonds.}
\vspace*{0.1cm}
\begin{tabular}{lcccccc}
 & $T [u_0/k_B]$ & $L_3/L$ & $c_{11}/c_{33}$ & $c_{44}/c_{33}$ & $\kappa$ & $\sqrt{c_{33}/4c_{44}}$ \\\hline
fl. & 0.01 & 5.5 & 0.026 & 0.038 & 2.8 & 2.6 \\
& 0.01 & 7.9 & 0.016 & 0.065 & 2.0 & 2.0 \\
& 0.1 & 5.8 & 0.030 & 0.041 & 2.7 & 2.5 \\
& 0.1 & 7.9 & 0.015 & 0.054 & 2.2 & 2.1 \\
\hline
sfl. & 0.01 & 3.4 & 0.12 & 0.10 & 1.7 & 1.6 \\
& 0.01 & 4.8 & 0.051 & 0.10 & 1.6 & 1.6 \\
& 0.1 & 3.4 & 0.087 & 0.086 & 1.9 & 1.7 \\
& 0.1 & 4.8 & 0.026 & 0.15 & 1.4 & 1.3\\
\end{tabular}
\end{table}
%****************************************************************


\begin{references}
\bibitem{dugdale-comm} Since $\lambda\gg 1$, this is equivalent to the
result of the Dugdale model, which describes the shape of the crazed
region assuming ideal plasticity \protect\cite{Wool1995}.
\end{references}
\end{document}